\documentclass[onecolumn]{aastex631}

\usepackage{amssymb}
\usepackage{amsmath}
\usepackage{gensymb}
\usepackage{graphicx}
\usepackage{xcolor}
\usepackage{natbib}

\newcommand{\hi}{H{\sc i}}

\newcommand{\mh}{\rm M_{H{\textsc i}}}
\newcommand{\macc}{\rm M_{acc}}
\newcommand{\mat}{\rm M_{atom}}
\newcommand{\mion}{\rm M_{ion}}
\newcommand{\mmol}{\rm M_{mol}}
\newcommand{\mstar}{\rm M_*}

\newcommand{\hii}{H{\sc i} 21\,cm}
\newcommand{\msun}{\rm M_{\odot}}

\shorttitle{Accretion rate of galaxies since $z \approx 0.35$}
\shortauthors{A. Bera et al.}
\graphicspath{{./}{figures/}}

\begin{document}

\title{The Gas Accretion Rate of Star-forming Galaxies over the last 4 Gyr}

\correspondingauthor{Nissim Kanekar}
\email{nkanekar@ncra.tifr.res.in}

\author{Apurba Bera}
\affil{International Centre for Radio Astronomy Research, Curtin University, Bentley, WA 6102, Australia}
\affil{Inter-University Centre for Astronomy and Astrophysics, Pune 411007, India}
\affil{National Centre for Radio Astrophysics, Tata Institute of Fundamental Research, Pune 411007, India}

\author{Nissim Kanekar}
\affil{National Centre for Radio Astrophysics, Tata Institute of Fundamental Research, Pune 411007, India}

\author{Jayaram N. Chengalur}
\affil{National Centre for Radio Astrophysics, Tata Institute of Fundamental Research, Pune 411007, India}

\author{Jasjeet S. Bagla}
\affil{Indian Institute of Science Education and Research Mohali, Knowledge City, Sector 81, Sahibzada Ajit Singh Nagar, Punjab 140306, India}

%---------------------------------------------------------------------

\begin{abstract}

Star-forming galaxies are believed to replenish their atomic gas reservoir, which is consumed in star-formation, through accretion of gas from their circumgalactic mediums (CGMs). However, there are few observational constraints today on the gas accretion rate in external galaxies. Here, we use our recent measurement of the scaling relation between the atomic hydrogen (H{\sc i}) mass $\rm{M_{HI}}$ and the stellar mass $\rm{M_*}$ in star-forming galaxies at $z\approx0.35$, with the relations between the star-formation rate (SFR) and $\rm{M_*}$, and the molecular gas mass $\rm{M_{Mol}}$ and $\rm{M_*}$, and the assumption that star-forming galaxies evolve along the main sequence, to determine the evolution of the neutral gas reservoir and the average net gas accretion rate onto the disks of star-forming galaxies over the past 4~Gyr. For galaxies with $\rm{M_*}\gtrsim 10^9\,{M_\odot}$ today, we find that both $\rm{M_*}$ and $\rm{M_{HI}}$ in the disk have increased, while $\rm{M_{Mol}}$ has decreased, since $z\approx0.35$. 
%The fractional change in the stellar mass (relative to the stellar mass today) is higher for lower-mass galaxies, while the opposite is true for the fractional changes in the molecular and atomic gas masses (each relative to its value today). The fractional change in the total baryonic mass, relative to the baryonic mass today, is $\approx40$\% for galaxies with stellar masses $\approx{10^9-10^{10.6}}\,\rm{M_\odot}$ today. 
The average gas accretion rate onto the disk over the past 4~Gyr is similar to the average SFR over this period, implying that main-sequence galaxies have maintained a stable H{\sc i} reservoir, despite the consumption of gas in star-formation. We obtain an average net gas accretion rate (over the past 4~Gyr) of ${\approx6\,\rm{M_\odot}}$~yr$^{-1}$ for galaxies with the stellar mass of the Milky Way. At low redshifts, $z \lesssim 0.4$, the reason for the decline in the cosmic SFR density thus appears to be the inefficiency in the conversion of atomic gas to molecular gas, rather than insufficient gas accretion from the CGM.
\end{abstract}


\section{Introduction} 
\label{sec:intro}

Neutral gas is the primary constituent of the interstellar medium (ISM) of star-forming galaxies. It provides the raw material for star-formation, and is consumed in the process. The gas reservoir is expected to be replenished  through accretion of gas from the circumgalactic medium (CGM) onto the `disks' of galaxies. The accretion may occur either through cooling of the hot virialized gas in the CGM \citep[the ``hot mode''; e.g.][]{rees77,white78} or through gas inflow along cold filaments \citep[the ``cold mode''; e.g.][]{binney77,birnboim03,keres05mnras}. 
%Cosmological simulations and semi-analytic models of galaxy evolution support the importance of gas inflows onto the disks of galaxies \citep[e.g.][]{correa18mnras, lagos14mnras, ho19apj, mitchel20mnras}. 
However, observational evidence for gas accretion in external galaxies has been scarce, partly because the inflowing gas is diffuse and difficult to detect, and partly due to the lack of unambiguous signatures of accretion. Indirect evidence of gas accretion onto galaxy disks has been found in several recent studies \citep[e.g.][]{cheung16apj,spring17mnras,kleiner17mnras, rahmani18mnras,zahedy19mnras}. Indeed, insufficient gas accretion to replenish the neutral gas reservoir of galaxies has been proposed to explain the observed decline in the cosmic star-formation rate (SFR) density at $z \lesssim 1$ \citep[e.g.][]{chowdhury20nature, chowdhury22survey,chowdhury22apjl}. 

\citet{scoville17apj} introduced an interesting approach to determine the gas accretion rate as a function of redshift, using dust continuum measurements to infer the ISM masses of galaxies (via an assumed dust-to-gas ratio) and then fitting for the dependence of the ISM mass on the galaxy redshift, stellar mass, and offset from the star-forming main sequence \citep[see also][]{scoville23apj}. They combined the above scaling relation with the assumption of the continuity of main-sequence evolution to infer the gas accretion rate in main-sequence galaxies. We note that the continuity of the main sequence is a standard assumption in the literature \citep[e.g.][]{renzini09,peng10apj, leitner12smg,speagle14,ciesla17ms}, with support from both observational evidence \citep[e.g.][]{rodighiero11apj} and hydrodynamical simulations \citep[e.g.][]{sparre2015ms, tacchella16ms}.
However, the dust-to-gas ratio is known to depend critically on galaxy metallicity; the calibration of the ISM mass is thus only applicable to massive galaxies, with near-solar metallicity \citep[stellar mass, $\rm M_* \gtrsim 2 \times 10^{10} \ \msun$ at high redshifts; ][]{scoville17apj}. Further, even for galaxies with near-solar metallicity in the central regions, the dust-to-gas ratio in the outer disk (which contains a significant fraction of the atomic phase) would be lower than in the central regions \citep{draine07}. As noted by \citet{scoville17apj, scoville23apj}, the inferred ISM masses are applicable to the inner disks of galaxies, where the assumption of near-solar metallicity is reasonable. The total ISM mass is hence likely to be under-estimated by this approach.

The atomic phase (made up of mainly atomic hydrogen, \hi, and helium) is known to dominate the neutral gas reservoir in main-sequence star-forming galaxies at $z \approx 0$, accounting for $\gtrsim 85\%$ of the neutral gas mass \citep[e.g.][]{saintonge17apjs,catinella18mnras}. 
%The accreted gas is also predominantly atomic in nature. 
Measurements of the dependence of the \hi\ mass of galaxies on the stellar mass and redshift would thus provide a more reliable way of determining the gas accretion rate, compared to the estimates of the total ISM mass. In the local Universe, the dependence of the \hi\ mass ($\mh$) on the stellar mass ($\rm M_*$) has been determined via studies of individual galaxies in the \hii\ line \citep[e.g.][]{catinella18mnras,parkash18apj}. Unfortunately, the weakness of this line, the main probe of the \hi\ mass of galaxies, has meant that it is very difficult to determine such \hi\ scaling relations at cosmological distances via studies of individual galaxies.

We have recently applied the technique of \hii\ stacking \citep{zwaan00thesis,chengalur01aa} to \hii\ data from a deep Giant Metrewave Radio Telescope (GMRT) survey of the Extended Groth Strip \citep[EGS; ][]{bera19apjl,bera22himf} to determine the $\rm \mh - M_*$ scaling relation for star-forming galaxies at $z \approx 0.35$ \citep[][see also \citet{sinigaglia22,chowdhury22scaling}]{bera23scaling}. In this {\it Letter}, we use this $\rm \mh - M_*$ relation, with the main-sequence relation, the scaling relation between molecular gas mass and stellar mass, and the assumption of the continuity of main-sequence evolution, to study the evolution of the different baryonic components of star-forming galaxies over the past 4~Gyr, and to determine the average gas accretion rate over this period.\footnote{Throughout this work, we use a flat $\Lambda$-cold dark matter ($\Lambda$CDM) cosmology, with ($\rm H_0$, $\rm \Omega_{m}$, $\rm \Omega_{\Lambda})=(70$~km~s$^{-1}$~Mpc$^{-1}$, $0.3, 0.7)$. Further, all stellar mass and SFR estimates assume a Chabrier initial mass function \citep{chabrier03imf}.}

\section{The evolution of the baryonic content of galaxies} 
\label{sec:baryons}

The baryonic mass of a galaxy is made up of stars, atomic gas, molecular gas, and ionized gas, with a small contribution from interstellar dust. 
While the stars and the neutral atomic and molecular gas are predominantly found in the disk of a galaxy, the ionized gas is found in both the disk and the CGM. The baryonic content of a galaxy increases due to accretion of gas from the CGM and the intergalactic medium, and can decrease due to gas outflows driven by supernovae or stellar winds. The net amount of gas accreted over a given time is the difference between the amount of gas accreted and the amount of gas lost in outflows; we will combine these effects to describe the evolution of the net accreted gas mass, $\macc$.
The change in the total baryonic content in the disk of a galaxy over a given time can then be written as 
\begin{equation}
\rm \macc = \Delta \mstar/{\it (1 - f_{return})} + \Delta \mmol + \Delta \mat + \Delta \mion 
\label{eqn:accretion}
\end{equation}
where $\Delta \mstar$, $\Delta \mat$, $\Delta \mmol$, and $\Delta \mion$ are the changes in, respectively, the stellar mass $\mstar$, the atomic gas mass $\mat \equiv 1.38\times \mh$, the molecular gas $\mmol$, and the ionized gas mass $\mion$, over this time. $\mh$ is the \hi\ mass and the factor of 1.38 in $\mat$ accounts for the contribution of helium. Finally, the factor $(1 - f_{return})$ accounts for the fraction of stellar mass that is returned to the gas phase \citep[see below; ][]{leitner11,scoville17apj}.

 In the above equation, the stellar mass of a star-forming galaxy is expected to increase with time, at the expense of the neutral gas mass, along with some mass loss due to stellar winds and supernovae. The molecular gas mass increases at the expense of the atomic gas mass, and decreases due to star-formation. The neutral atomic gas mass decreases via conversion to molecular gas, but increases through accretion onto the disk. Finally, the ionized gas mass decreases due to conversion to the neutral atomic phase, but increases due to stellar and supernova-driven outflows. We will neglect the ionized gas mass in what follows, as its mass in the disk is expected to be much lower than the neutral gas mass \citep[e.g.][]{draine11book}.

Using the main-sequence relation and its redshift evolution, and the $\mh - \mstar$ and $\mmol - \mstar$ scaling relations at any pair of redshifts, we can determine the changes in the stellar mass, the neutral atomic gas mass, and the neutral molecular gas mass of galaxies over the redshift range, as a function of their stellar mass. Finally, substituting for $\Delta \mstar$, $\Delta \mmol$, and $\Delta \mat$ in Equation~\ref{eqn:accretion} would yield the net gas mass accreted by galaxies between the two redshifts, and thus the average net gas accretion rate. We will apply this formalism to the redshift range $z \approx 0-0.35$, to determine the net average gas accretion rate over the last 4~Gyr.

\subsection{Stellar mass build-up along the main sequence} 
\label{sec:stellarmass}

Star-forming galaxies are known to show a tight correlation between the SFR and $\mstar$, known as the main sequence \citep[e.g.][]{madau14ara}. The main-sequence relation has been shown to exist out to $z \approx 6$ \citep[e.g.][]{popesso23}, and is known to evolve with redshift; the evolution has been described using various parametric forms \citep[e.g.][]{whitaker12apj, whitaker14apj, lee15apj, leslie20apj, popesso23}. Star-forming field galaxies are thought to evolve along the main sequence, i.e. as their stellar mass increases, their SFR changes accordingly to keep them on the main sequence. The knowledge of the redshift evolution of the main-sequence relation can hence be used to trace the stellar-mass history of present-day main-sequence galaxies over their lifetimes \citep[see, e.g.,][]{renzini09,peng10apj,leitner12,speagle14,scoville17apj, scoville23apj}.  

Following \citet{scoville17apj,scoville23apj}, we will restrict ourselves to main-sequence galaxies and assume the principle of continuity of main-sequence evolution. We will ignore both major mergers\footnote{The rate of major mergers is not significant for star-forming galaxies in the stellar mass range considered in this work \citep[e.g.][]{rodriguezgomez15merger}; the stellar mass growth is dominated by star-formation for these galaxies \citep{guo08mnras}.} (which can remove galaxies from the main sequence) and the quenching of star-formation activity. 
Assuming that today's main-sequence galaxies were also on the main sequence 4~Gyr ago, i.e. at $z \approx 0.35$, we can use the redshift-dependent main-sequence relation to estimate the net change in their stellar masses from $z \approx 0.35$ to $z \approx 0$. The increase in the stellar mass of a main-sequence galaxy from $z \approx 0.35$ to the present epoch is given by
\begin{equation}
{\rm \Delta \mstar \equiv M_{*,0} - M_{*,0.35}} = (1-f_{return}) {\rm \int_{{\it z}=0.35}^{0} {\rm SFR} (M_*,} z) \:dz
\label{eqn:mstar}
\end{equation} 
where $\rm M_{*,0.35}$ and $\rm M_{*,0}$ are the initial (at $z \approx 0.35$) and final (at $z=0$) stellar masses of the galaxy, respectively, $\rm SFR(M_*,z)$ is the redshift-dependent main-sequence relation, and $f_{return}$ is the fraction of stellar mass that is returned to the gas phase via stellar winds or supernovae. The value of $f_{return}$ depends on the initial mass function and typically lies in the range $0.27 - 0.41$ \citep[see, e.g.,][]{madau14ara}. Here, we assume $f_{return} = 0.3$, applicable for a Chabrier initial mass function \citep{leitner11}\footnote{Our results do not change significantly if a different value of $f_{return}$, within the range $0.27 - 0.41$, is assumed.}. We also assume that this processed gas is not available for further star-formation \citep[e.g.][]{scoville17apj}. We use the redshift-dependent main-sequence relation of \citet{whitaker12apj}\footnote{Errors associated with the main-sequence relation have been ignored in this work. Uncertainties in the scaling relations dominate the total errors in our final results.},
\begin{equation}
{\rm log [SFR]} = \alpha (z) {\rm \left[ log (M_*/M_\odot) - 10.5 \right]} + \beta (z) \, ,
\label{eqn:whitaker12}
\end{equation}
where $\alpha (z) = 0.70 - 0.13 z$ and $\beta (z) = 0.38 + 1.14 z - 0.19 z^2$, to determine the stellar mass at $z \approx 0.35$ of main-sequence galaxies with present-day stellar mass $\rm M_{*,0}$. We restrict to galaxies with $\rm M_{*,0} \geq 10^9 \; M_{\odot}$, for which the local $\mh - \mstar$ scaling relation has been robustly measured today \citep[e.g.][]{catinella18mnras, parkash18apj}.
Using Equation~\ref{eqn:mstar}, this corresponds to galaxies with stellar masses $\gtrsim 10^{8.5} \ \msun$ at $z \approx 0.35$.

\begin{figure*}
\begin{center}
\includegraphics[scale=0.705]{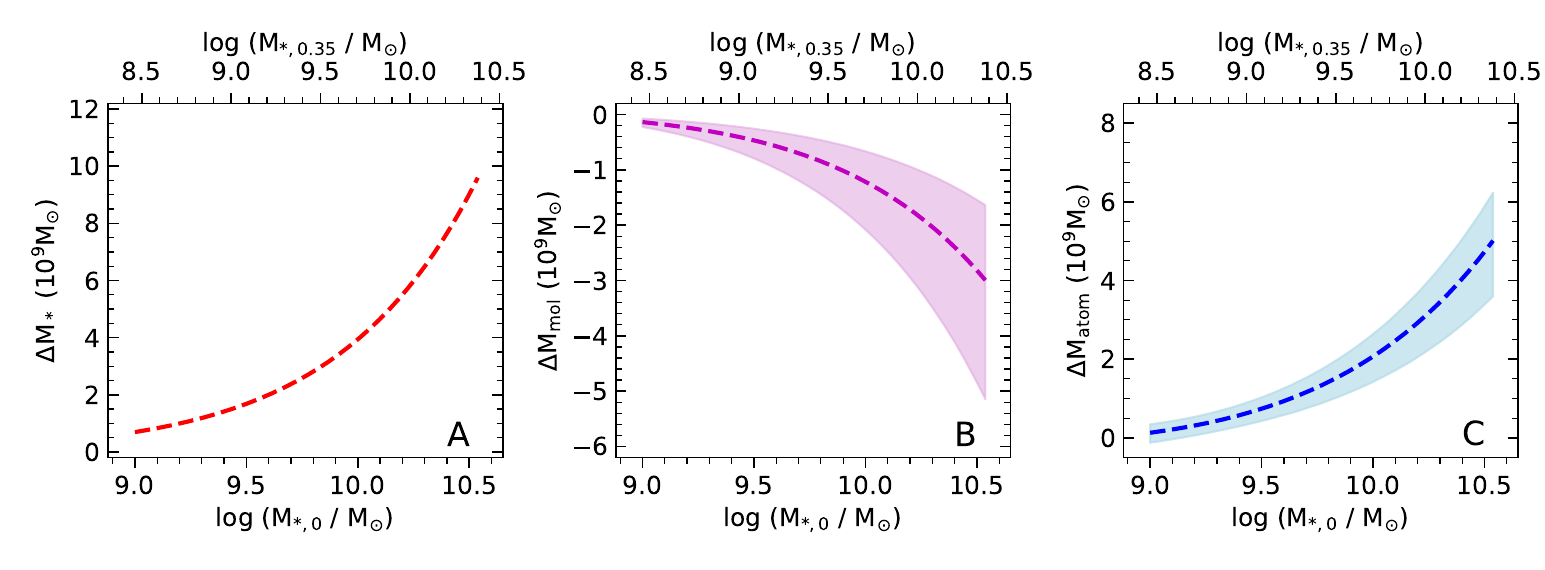}
\caption{Net changes in the average [A]~stellar mass ($\rm M_*$), [B]~molecular gas mass ($\rm M_{\rm mol}$), and [C]~atomic gas mass ($\rm M_{\rm atom}$) in the disks of main-sequence galaxies from $z \sim 0.35$ to $z \sim 0$ are shown as functions of their present-day stellar mass ($\rm M_{*,0}$, bottom axis) and their initial stellar mass ($\rm M_{*,0.35}$, top axis). The shaded regions show the 68\% confidence intervals.
%In panel [C], the change in the atomic gas mass was estimated using two different measurements of the ${\rm \mh - M_*}$ scaling relation at $z\approx 0.35$, from the uGMRT \hii~survey of the EGS \citep[solid blue line;][]{bera23scaling} and the MIGHTEE-HI survey \citep[dotted red line;][]{sinigaglia22}.
}
\label{fig:evolution}
\end{center}
\end{figure*}

\begin{figure*}
\begin{center}
\includegraphics[scale=0.705]{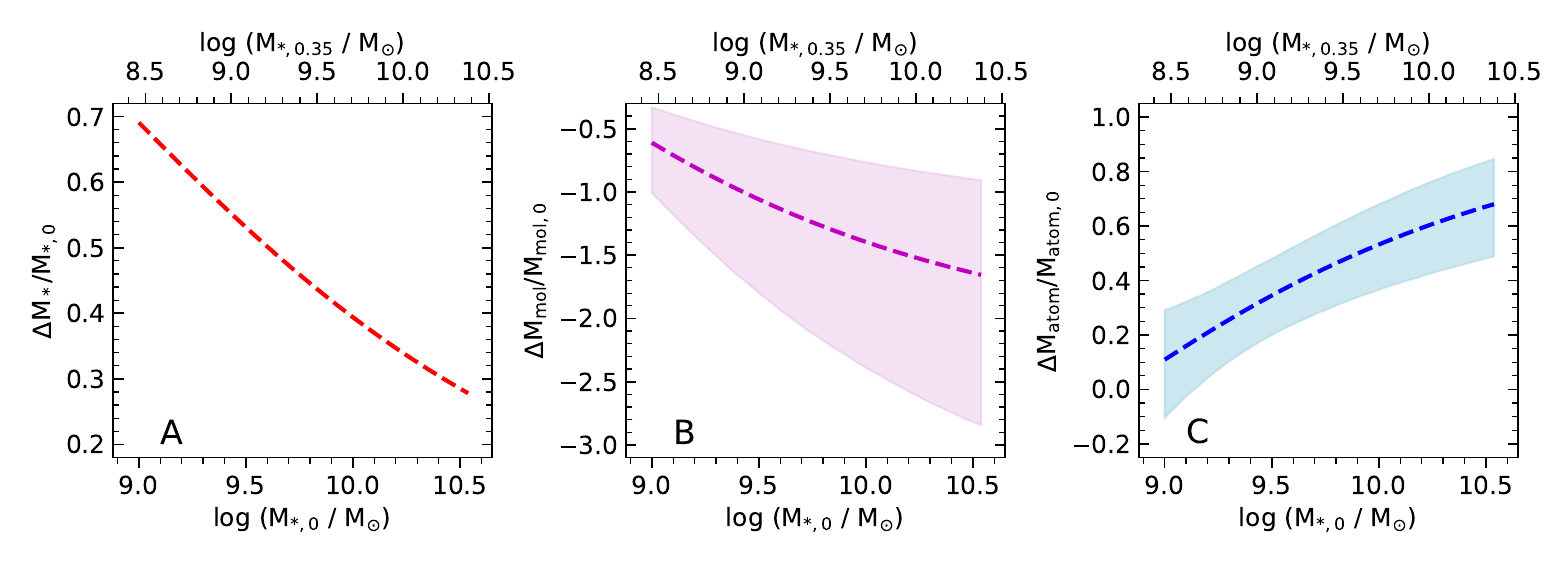}
\caption{Fractional changes, with respect to their respective present day values, in the average [A]~stellar mass ($\rm \Delta M_*/M_{*,0}$), [B]~molecular gas mass ($\rm \Delta M_{\rm mol}/M_{\rm mol,0}$), and [C]~atomic gas mass ($\rm \Delta M_{\rm atom}/M_{\rm atom,0}$) in the disks of main sequence galaxies from $z \sim 0.35$ to $z \sim 0$ are shown as functions of their present-day stellar mass ($\rm M_{*,0}$, bottom axis) and their initial stellar mass ($\rm M_{*,0.35}$, top axis). The shaded regions show the 68\% confidence intervals.
%In panel [C], the change in the atomic gas mass was estimated using two different measurements of the ${\rm \mh - M_*}$ scaling relation at $z\approx 0.35$, from the uGMRT \hii~survey of the EGS \citep[solid blue line;][]{bera23scaling} and the MIGHTEE-HI survey \citep[dotted red line;][]{sinigaglia22}.
}
\label{fig:evolution_frac}
\end{center}
\end{figure*}

\subsection{Molecular gas scaling relations} 
\label{subsec:molecular}

The evolution of the molecular gas content of main-sequence galaxies has been quantified through measurements of the redshift-dependent scaling relation between $\mmol$ and $\mstar$ \citep[e.g.][]{genzel15apj,tacconi18apj}. The molecular gas mass of galaxies is typically estimated from the CO rotational lines, the far-infrared dust continuum emission, or the $\approx 1$~mm dust continuum  \citep[see][for a review]{tacconi20araa}. We used the redshift-dependent scaling relation\footnote{This relation includes the contribution of helium \citep{tacconi20araa}.} connecting the ratio of the molecular gas mass to the stellar mass, $\rm \mu_{Mol} \equiv \left[\mmol/\mstar \right]$ of a galaxy to its stellar mass \citep{tacconi20araa}, 
\begin{equation}
{\rm log \left[\mu_{\rm Mol}\right]} = A + B\:[{\rm log}(1+z) - F]^2 + D[{\rm log(M_*/M_{\odot})}- 10.7]  
\label{eqn:molecules}
\end{equation}
where $A = 0.06 \pm 0.20$, $B = -3.3 \pm 0.2$,  $D = -0.41 \pm 0.03$, and $F=0.65 \pm 0.05$, to determine the molecular gas mass of a main-sequence galaxy from its stellar mass.\footnote{Note that we assume that the offset of each galaxy from the main sequence is zero.} For each galaxy with present-day stellar mass $\rm M_{*,0}$, we can combine Equations~\ref{eqn:whitaker12} and \ref{eqn:molecules} to determine its molecular gas mass at $z = 0$ and $z = 0.35$, and thus  estimate the change in its molecular gas mass $\Delta \mmol$ between $z = 0.35$ and $z = 0$ from the relation
\begin{equation}
\rm \Delta \mmol = M_{Mol, 0} - M_{Mol, 0.35} = \mu_{Mol, 0} \; M_{*,0} - \mu_{Mol, 0.35} \; M_{*, 0.35}
\label{eqn:dmol}
\end{equation} 
where $\rm M_{*,0.35}$ and $\rm M_{*,0}$ are again the initial (at $z=0.35$) and final (at $z=0$) stellar masses, respectively, and $\mu_{{\rm Mol}, z}$ can be inferred from Equation~\ref{eqn:molecules}.

We note that the galaxies used to measure the scaling relation parameters have stellar masses in the range $\mstar = 10^{9} - 10^{12.2} \ \msun$ \citep{tacconi20araa}. We assume that the same scaling relation is also applicable to lower-mass galaxies, with $\mstar \approx 10^{8.5} \ \msun$, which are part of the EGS sample at $z \approx 0.35$ \citep{bera23scaling}.

\subsection{Atomic gas scaling relations} 
\label{subsec:atomic}

For neutral atomic gas, the $\mh - \mstar$ scaling relation is known at $z \approx 0$ from direct \hii\ emission studies of individual galaxies \citep[e.g.][]{catinella18mnras,parkash18apj}. We will use the $\mh - \mstar$ relation obtained for blue, star-forming galaxies of the xGASS sample \citep{catinella18mnras,bera23scaling}\footnote{Note that using the  $\mh - \mstar$ scaling relation of \citet{parkash18apj} does not significantly change our results.}
\begin{equation}
{\rm log}(\mh/M_{\odot}) = (8.934 \pm 0.036) + (0.516 \pm 0.030)\left[ {\rm log}(M_*/M_{\odot}) - 9.0\right] \,.
\label{eqn:xgass}
\end{equation}

At present, there are no estimates of the $\mh - \mstar$ relation at cosmological distances, $z \gtrsim 0.1$, based on \hii\ studies of individual galaxies. However, we have recently used the GMRT to carry out a deep \hii\ emission survey of the EGS, which yielded an estimate of the ``mean'' $\mh - \mstar$ relation in blue, star-forming galaxies at $z \approx 0.35$ with $\mstar = 10^{8.0} - 10^{10.4} \ \msun$, based on stacking the \hii\ emission from galaxies in different stellar-mass bins \citep{bera23scaling}.
%\footnote{We note that an independent estimate of the $\mh - \mstar$ relation at $z \approx 0.37$ from a similar \hii\ stacking  analysis was obtained by \citet{sinigaglia22}, although their results are affected by both clear systematic effects and low-significance detections in some of the stacked \hii\ spectra \citep{bera23scaling}. We have hence not used this scaling relation in the present work, but note that using this $\mh - \mstar$ relation in place of Equation~\ref{eqn:hi-egs} yields consistent results, within $\lesssim 2.5\sigma$ significance over the entire stellar mass range.}. 
As noted by \citet{bera23scaling}, this mean relation may be combined with an assumed lognormal scatter in the $\mh - \mstar$ relation to infer the ``median'' $\mh - \mstar$ scaling relation, which can be directly compared to the scaling relation obtained from a fit to \hii\ emission measurements in individual galaxies. Assuming a lognormal scatter in the $\mh - \mstar$ relation at $z \approx 0.35$ that is equal to that at $z \approx 0$ in the xGASS sample, \citet{bera23scaling} give the following median $\mh - \mstar$ scaling relation for blue, star-forming galaxies at $z \approx 0.35$ in the EGS, 
\begin{equation}
{\rm log(\mh/\msun) = (8.977 \pm 0.069) + (0.183 \pm 0.104)\left[ log(\mstar/\msun) - 9.0 \right]} \, .
\label{eqn:hi-egs}
\end{equation}

The net change in the atomic gas mass of a main-sequence galaxy from $z = 0.35$ to $z = 0$ can then be estimated using the relation
\begin{equation}
\rm \Delta \mat  = 1.38 \left[ M_{\textrm{\hi}, 0}(M_{*,0}) - M_{\textrm{\hi}, 0.35}(M_{*, 0.35}) \right]  \, ,
\label{eqn:hi}
\end{equation} 
where $\rm M_{\textrm{\hi}, 0.35}$ and $\rm M_{\textrm{\hi}, 0}$ are the initial (at $z = 0.35$) and final (at $z=0$) \hi\ masses, respectively, $\rm M_{*, 0.35}$ and $\rm M_{*, 0}$ are the initial and final stellar masses respectively, and $\textrm{M}_{\textrm{\hi}, z}(\textrm{M}_{*, z})$ is the $\mh - \mstar$ scaling relation (i.e. Equation~\ref{eqn:hi-egs}) at the redshift $z$.

\section{The evolution of the baryonic composition of galaxies from $z \approx 0.35$} 
\label{sec:infallrate}

We have used Equations~\ref{eqn:mstar}--\ref{eqn:hi}
%\ref{eqn:whitaker12}, \ref{eqn:molecules}, \ref{eqn:xgass}, \ref{eqn:hi-egs}, and 
 to determine the changes in the stellar mass, the molecular gas mass, and the atomic gas mass in the disks of present-day main-sequence galaxies between $z \approx 0.35$ and $z = 0$. As noted earlier, we assume continuity of main-sequence evolution, i.e. that the galaxies evolve along the main sequence \citep{scoville17apj,scoville23apj}.

Figures~\ref{fig:evolution}[A--C] show, respectively, the changes in the stellar mass $\Delta \mstar$, the molecular gas mass $\Delta \mmol$, and the atomic gas mass $\Delta \mat$, over the redshift range $z = 0.35$ to $z = 0$, as a function of the stellar mass of galaxies today, $\rm M_{*,0}$. Figures~\ref{fig:evolution_frac}[A--C] show, respectively, the fractional changes in the above three quantities (relative to their present-day values) over the same period as a function of the stellar mass today.  It is clear (see Figures~\ref{fig:evolution}[A] and \ref{fig:evolution_frac}[A]) that the stellar masses of today's main-sequence galaxies have increased significantly over the past 4~Gyr. Relatively low-mass galaxies, with $\rm M_{*,0} \approx 10^9\ \msun$, have acquired $\approx 70$\% of their current stellar mass since $z \approx 0.35$, while high-mass galaxies, with $\rm M_{*,0} \approx 10^{10.5} \ \msun$, have acquired $\approx 30$\% of their present stellar mass during this period.

Conversely, Figure~\ref{fig:evolution}[B] shows that the molecular gas content of all galaxies has declined significantly over the last 4~Gyr. The fractional decline relative to the present-day molecular gas mass is seen (in  Figure~\ref{fig:evolution_frac}[B]) to be the highest for the highest-mass galaxies, with $\Delta \mmol/\rm M_{mol,0} \lesssim -1$ for $\rm M_{*,0} \gtrsim 10^{9.5}\ \msun$. In other words, the molecular gas reservoir of present-day main-sequence galaxies has been steadily consumed by star-formation activity since $z \approx 0.35$.

Finally, the solid blue curves (and blue shaded regions) in Figures~\ref{fig:evolution}[C] and \ref{fig:evolution_frac}[C] show the evolution of the atomic gas mass and the fractional atomic gas mass, relative to today's atomic gas mass, of galaxies over the redshift range $z \approx 0.35-0$.  We find that $\Delta \mat$ is always positive, implying a net increase in the atomic gas mass of galaxies over the last 4~Gyr for all stellar masses. The fractional change in the atomic gas mass (relative to the atomic gas mass today) is low ($\approx 10$\%) for low-stellar-mass galaxies (with $\rm M_{*,0} \approx 10^{9}\ \msun$), but substantial ($\approx 70$\%) for the highest-stellar-mass galaxies today.

\begin{figure*}[h!]
\begin{center}
\includegraphics[scale=0.75]{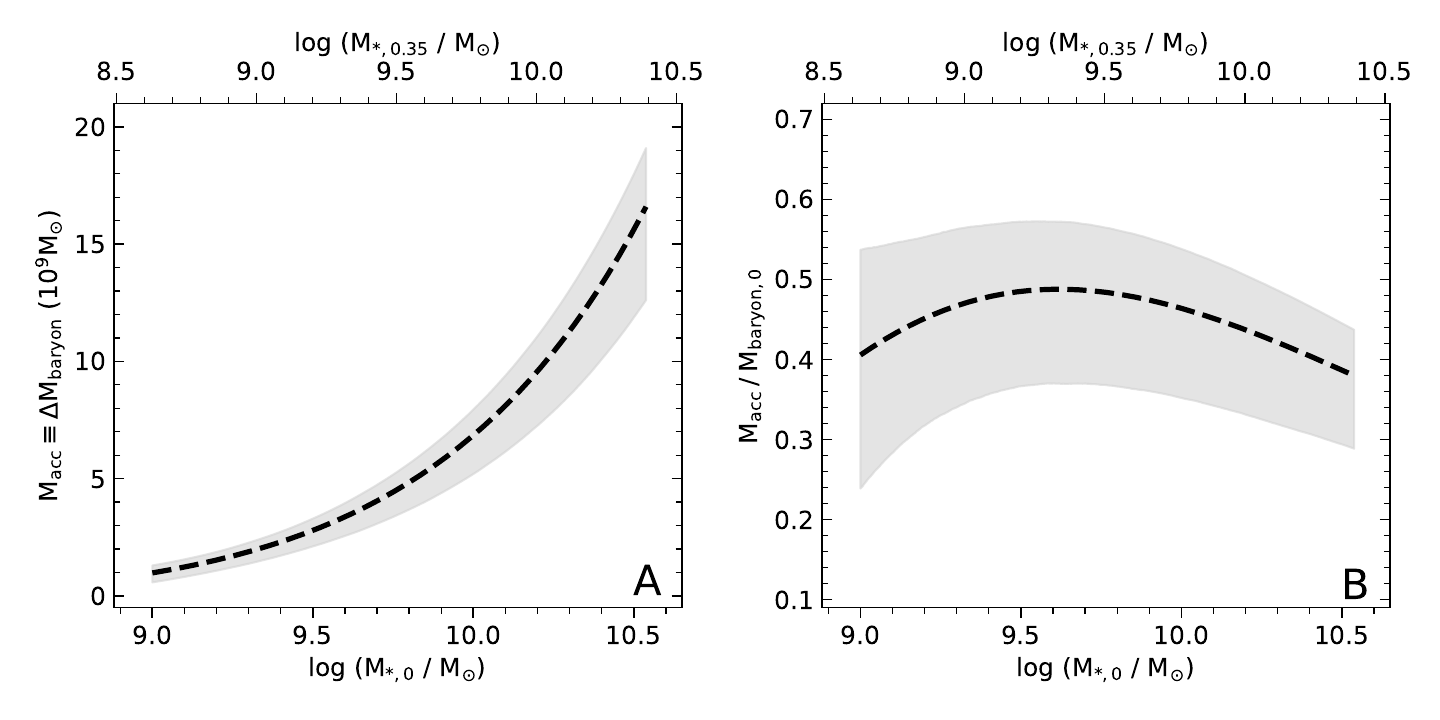}
\caption{[A]~Net change in the average baryonic mass ($\rm \Delta M_{baryon} \equiv M_{acc}$) in the disks of main-sequence galaxies from $z \sim 0.35$ to $z \sim 0$ as a function of their present-day stellar mass ($\rm M_{*,0}$, bottom axis) and their initial stellar mass ($\rm M_{*,0.35}$, top axis). 
[B]~The fractional change in the average baryonic mass, with respect to the present-day baryonic mass, as a function of the present-day stellar mass (bottom axis) and the initial stellar mass (top axis). The shaded regions in the figures show the 68\% confidence intervals for the corresponding curves.}
\label{fig:evolution_baryon}
\end{center}
\end{figure*}

Figure~\ref{fig:evolution_baryon}[A] plots the change in the total baryonic mass $\rm \Delta M_{baryon}$ of main-sequence galaxies from $z \approx 0.35$ to $z = 0$ against their stellar mass today, $\rm M_{*,0}$. We note that $\rm \Delta M_{baryon} \equiv M_{acc}$, the net average gas mass accreted over the last $\approx 4$~Gyr (i.e. the difference between the gas  mass accreted and the gas mass lost due to winds or outflows). The figure shows that all galaxies with stellar mass in the range $\approx 10^9 - 10^{10.6} \, \msun$ today have increased their total baryonic mass over the last four Gyr, with the increase being larger for higher stellar masses. However, the fractional change in the total baryonic mass, relative to the baryonic mass today (i.e. $\rm M_{acc}/M_{baryon,0}$), is seen in Fig.~\ref{fig:evolution_baryon}[B] to be approximately constant, ${\approx 40}$\%, across the above stellar mass range.

\begin{figure*}
\begin{center}
\includegraphics[scale=0.72]{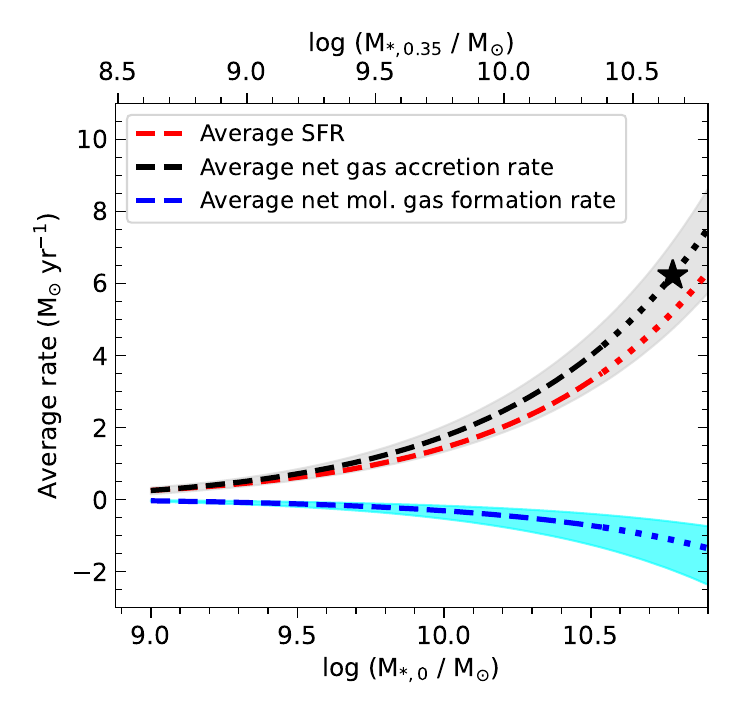}
\caption{The time-averaged SFR (dashed red curve), the average net molecular gas formation rate (dashed blue curve) and the average net gas accretion rate of the disks of present-day main sequence galaxies (dashed black curve) over the past 4~Gyr are shown as functions of their present-day stellar mass ($\rm M_{*,0}$, bottom axis) and their initial stellar mass ($\rm M_{*,0.35}$, top axis). The dotted curves show the results for the average rates on extrapolating the \hi\ scaling relation of \citet{bera23scaling} to a stellar mass of $\mstar = 10^{10.7}\, \msun$ at $z \approx 0.35$. The star indicates the average net gas accretion rate for galaxies with stellar masses equal to that of the Milky Way.
The shaded regions show the 68\% confidence intervals for the average net gas accretion rate and the average net molecular gas formation rate.}
\label{fig:accretionrate}
\end{center}
\end{figure*}

For a galaxy of a given stellar mass today, the ratio of $\rm M_{acc}$ to the elapsed time $\Delta t$ between any two redshifts gives the time-averaged net gas accretion rate $\rm \dot{\langle M \rangle}_{acc}$ between the two redshifts. Fig.~\ref{fig:accretionrate} plots (dashed black curve) the above average net gas accretion rate from $z \approx 0.35$ to $z \approx 0$ (i.e. over the last $\approx 4$~Gyr) as a function of the stellar mass of galaxies today, $\rm M_{*,0}$; the grey shaded region shows the 68\% confidence interval. The dashed red curve shows the average SFR ($\equiv \Delta \mstar/ \left[ (1-f_{return}) \Delta t \right]$) over the same period, again as a function of the stellar mass today, while the dashed blue curve and blue shaded region show the average net formation rate of molecular hydrogen (i.e. the difference between the formation rate and the destruction rate). The red and black curves are in good agreement, within the errors: this indicates that the average rate of net accretion of gas onto main-sequence galaxies over the last $\approx 4$~Gyr is sufficient to balance the average SFR in these galaxies. Star-forming galaxies on the main sequence, with $\rm M_* \approx 10^9 - 10^{10.6}\, \msun$ today, have thus accreted substantial amounts of gas over the last $\approx 4$~Gyr to replenish their neutral gas reservoir, maintaining a stable (indeed, slightly increasing) \hi\ reservoir, despite the continuous consumption of \hi\ in the star-formation process. We emphasize that this is very unlike the situation towards the end of the epoch of galaxy assembly, $z \approx 1$, where \citet{chowdhury22apjl} find evidence that insufficient gas accretion is the cause of the decline in the SFR density at $z < 1$.

Conversely, it is clear from Fig.~\ref{fig:accretionrate} that the average net rate of H$_2$ formation over $z \approx 0 - 0.35$ is significantly lower than both the average SFR and the average gas accretion rate. Indeed, the net rate of molecular gas formation is negative, indicating that the atomic-to-molecular gas conversion does not keep pace with the conversion of molecular gas to stars. This implies that it is the inefficient conversion of atomic hydrogen to molecular hydrogen that is likely to be the main cause of the decline in the cosmic SFR density over the last 4~Gyr.

We note that there could be an environmental dependence to the various scaling relations \citep[e.g.][]{cortese11mnras,catinella13mnras}. The main-sequence and $\mmol - \mstar$ scaling relations are predominantly based on field galaxies \citep[e.g.][]{whitaker12apj,tacconi20araa}. We have used the Sloan Digital Sky Survey-DR8 group catalog of \citet{tempel12} to find that roughly half of the blue xGASS galaxies used to determine the $\mh - \mstar$ relation at $z \approx 0$ are field galaxies  \citep[see also][]{catinella13mnras}. We note that this estimate of $\approx 50$\% of the xGASS galaxies being field objects is likely to be a lower limit as many of the group galaxies are in ``groups'' with only $2-3$ members, and may thus well be field galaxies \citep{tempel12}. Similarly, \citet{gerke12apj} have used the Voronoi-Delaunay group finder to classify DEEP2 galaxies: of the 260 EGS galaxies used to determine the $\mh-\mstar$ relation at $z \approx 0.35$ and that have been classified by \citet{gerke12apj}, $\approx 70$\% are field galaxies (and a significant number of the remaining systems are in groups with $2-3$ members). It thus appears unlikely that the environmental dependence of the scaling relations would significantly affect our results.

The stellar mass of the Milky Way today is $\rm M_{*, 0} = (6.08 \pm 1.14) \times 10^{10} \, \msun$ \citep{licquia15}. This lies beyond the stellar mass range at $z = 0$ ($\rm M_{*,0} \approx 10^9 - 10^{10.6} \, \msun$) covered by our results. Assuming that we can extrapolate the \hi\ scaling relation at $z \approx 0.35$ to a stellar mass of $\mstar = 10^{10.7}\, \msun$, we can estimate the average net gas accretion rate of Milky Way-like galaxies over the last $\approx 4$~Gyr. The results are shown as the dotted curve in Fig.~\ref{fig:accretionrate}. We obtain an average net gas accretion rate of $\rm \approx 6 \, \msun$~yr$^{-1}$ (indicated by the star in the figure), over the past 4~Gyr for main-sequence galaxies with the stellar mass of the Milky Way. This is broadly consistent with estimates of the total gas accretion rate onto the Milky Way \citep[e.g.][]{fox14,richter17}. 

In passing, as noted by \citet{bera23scaling}, we emphasize that the $\mh - \mstar$ relation used here is based on a relatively small number of galaxies, and a small cosmic volume, and could hence well be affected by cosmic variance. The possibility of cosmic variance in this relation would affect the present results as well. A wide-field determination of the $\mh - \mstar$ relation at intermediate redshifts would allow a better estimate of the average gas accretion rate onto the disks of galaxies, using the approach described here.

\section{Summary} 
\label{sec:summary}

We present a formalism to determine the evolution of the baryonic composition of star-forming galaxies between any two redshifts, based on the main-sequence relation between SFR and stellar mass, the scaling relation between molecular gas mass and stellar mass, the scaling relation between atomic gas mass and stellar mass, and the assumption that star-forming galaxies continuously evolve along the main sequence. We apply this formalism to our recent estimate of the $\mh - \mstar$ relation at $z \approx 0.35$, to determine the average changes in the stellar, molecular gas, and atomic gas contents of the disks of star-forming galaxies from $z \approx 0.35$ to $z = 0$, as a function of the galaxy stellar mass today, for the stellar mass range $\rm M_{*,0} = 10^{9.0} - 10^{10.6} \, \msun$. We find that the stellar and atomic gas masses of today's main-sequence galaxies have both increased since $z \approx 0.35$, while the molecular gas masses of these galaxies have declined over the same period. The fractional increase in the stellar masses (relative to the present-day stellar mass) is $\approx 30-70$\% , with larger fractional increases at lower stellar masses, while the fractional increase in the atomic gas masses (relative to the present-day atomic gas mass) is $\approx 10-70$\%, with a larger fractional increase at high atomic gas masses. 

We combine the changes in the stellar mass, the molecular gas mass, and the atomic gas mass to determine the net change in the baryonic mass of main-sequence galaxies over the last 4~Gyr. We find that the fractional net increase in the baryonic mass of these galaxies, relative to the present-day baryonic mass, is $\approx 40$\% for stellar  masses today of $\rm \approx 10^9 - 10^{10.6} \, \msun$. Finally, we determine the average net gas accretion rate of star-forming galaxies over the last 4~Gyr, finding average net accretion rates of $\approx 0.2 - 5 \, \msun$~yr$^{-1}$, similar to the average SFR over this period. The average net gas accretion rate for Milky Way-like galaxies is ${\approx 6\, \msun}$~yr$^{-1}$ since $z \approx 0.35$. We thus find that main-sequence galaxies accrete sufficient amounts of gas over the last $\approx 4$~Gyr to maintain a stable (and slightly increasing) \hi\ reservoir, with the gas accretion compensating for the gas consumption via star-formation. The observed decline in the cosmic SFR density over the last $\approx 4$~Gyr thus appears to arise due to the inefficient conversion from \hi\ to H$_2$, which does not sufficiently replenish the amount of molecular gas consumed in the process of star-formation.

\section*{Acknowledgments}
%\begin{acknowledgments}
We thank the staff of the GMRT who have made these observations possible. The GMRT is run by the National Centre for Radio Astrophysics of the Tata Institute of Fundamental Research. AB and NK thank Aditya Chowdhury for many discussions on \hii\ stacking that have contributed to this paper. We also thank an anonymous referee whose detailed comments on an earlier version of the manuscript improved the paper. NK acknowledges support from the Department of Science and Technology via a Swarnajayanti Fellowship (DST/SJF/PSA-01/2012-13). AB, NK, $\&$ JNC also acknowledge the Department of Atomic Energy for funding support, under project 12-R\&D-TFR-5.02-0700. 
%\end{acknowledgments}

\bibliography{galaxyrefs}{}
\bibliographystyle{aasjournal}

\end{document}